# Fuzzy-Controlled Scheduling of Route-Request Packets (FSRR) in Mobile Ad Hoc Networks

**Abu Sufian[1], Anuradha Banerjee[2*] and Paramartha Dutta[3]**

[1]University of Gour Banga, Malda - 732103, West Bengal, India; sufian.asu@gmail.com
[2]Kalyani Government Engineering College, Kalyani - 741235, West Bengal, India; anuradha79bn@gmail.com
[3]Visva-Bharati University, Santiniketan, Bolpur - 731235, West Bengal, India; paramartha.dutta@gmail.com

## Abstract

In ad hoc networks, the scheduling of route-request packets should be different from that of message packets, because during transmission of message packets the location of the destination is known whereas in route discovery this is not known in most of the cases. The router has to depend upon the last known location, if any, of the destination to determine the center and radius of the circle that embeds all possible current position of the destination. Route-request packets generated from the source are directed towards this circle i.e., directional route discovery can be applied. Otherwise, when no earlier location of the destination is known the route-requested has to be broadcast in the whole network consuming a significant amount of time than directional route discovery. The present article proposes a fuzzy controlled scheduling of route-request packets in particular that greatly reduces the average delay in route discovery in ad hoc networks.

**Keywords:** Ad Hoc Networks, Delay, Embedding Circle, Route-Request, Scheduling

## 1. Introduction

A mobile ad hoc network a collection of wireless nodes which form a temporary network without relying on an existing infrastructure or centralized administration. These networks are deployed mainly in emergency situations like battlefield, natural disasters like earthquake, floods etc[1-5]. Many routing protocols have been proposed in ad hoc networks so far. In all of them, when the destination node is out of the radio-range of the source node the communication has to be multi-hop where some nodes act as router to bridge the gap between the source and destination nodes[6–10]. If a node receives multiple message forwarding requests, it serves one of them and stores the others in its queue. The order in which these requests will be served is termed as a schedule. The job of a scheduler is to pick up that particular schedule that is expected to produce the best performance.

The choice of scheduling algorithm has a significant effect on the overall performance of the route, especially when the traffic load is high[10–14]. There are different scheduling policies for different network scenarios. Different routing protocols use different methods of scheduling. Among them, FCFS (First-Come-First-Served) is quite heavily used. The drop-tail policy is used as a queue management algorithm in various scheduling algorithms for buffer management[13]. Except for the no-priority scheduling algorithm, all other scheduling algorithms give higher priority to control packets than to data packets. In no priority scheduling algorithm both control and data packets are served in FIFO (First-In-First-Out) order. Certain scheduling schemes depend on the size of the message and number of hops to traverse. In Smallest Message First (SMF)[11] algorithm, the packets are scheduled in ascending order of the size of messages of which they are a part.

In Smallest Remaining Message First Scheme (SRMF)[12,13] packets are ordered on the basis of the amount of message packets remaining to be sent after the current packet. On the other hand, in Shortest Hop Length First (SHLF) scheduling[11,14] the distance between the source and destination, measured in terms of the

** Author for correspondence*

number of hops, influences the time a packet needs to reach its destination. The packet with the shortest hop is assigned highest priority. The scheduling decision is made independently at each node. Some other priority schedulers[11,14] concentrate on packet delivery ratio. If a node $n_j$ produces higher packet delivery ratio for a node $n_i$ compared to another node $n_k$, then a packet from $n_j$ will be assigned higher priority in message queue of $n_i$ compared to a packet from $n_k$. Age of a packet in also considered as an important parameter to avert starvation in some priority scheduling schemes. Energy efficient schedulers also consider[11] residual energy of source and destination nodes.

To the best of the author's knowledge, as far as the scheduling of route-request packets is concerned, there does not exist any scheme in the literature for ad hoc networks. The present article focuses on proposing one scheduling scheme that will prioritize the route-requests with recent known locations of the destinations where the destination is close to the current router and embedding circle is small. When these criteria are satisfied, the current router takes much lesser time to play its part in route discovery since route-request packet does not need to be broadcast anymore; directional route discovery can be applied efficiently. Much lesser number of downlink neighbors of the current router is forwarded the route request by the current router compared to all the downlink neighbors of the current router as in the case with route request broadcast. This reduces the average waiting time of all nodes that initiated route discovery because route-requests known to be lesser costly are served first. The idea is based on the Shortest-Job-First (SJF) scheduling algorithm where the process with smallest burst time gets the CPU first.

## 2. FSRR in Detail

### 2.1 Determination of the Destination Embedding Circle (DEC)

The technique of determining DEC is based on the following assumptions:-

- $n_s$ is the source and $n_d$ is the destination node.
- $n_s$ initiates route discovery at time ts which arrives at a router $n_i$ at time t.
- The maximum lifetime of a RREQ packet is $\tau$.
- A router $n_i$ knows the location of the destination at time t1 where t1 < t; t being the current time.
- Location of any node $n_i$ in the network at time t′ is denoted by an ordered pair $(x_i(t'), y_i(t'))$.
- Maximum velocity of any node $n_d$ is given by $v_{max}(d)$.
- Approximate velocity of the wireless signal is given by $v_s$

The circle that embeds all possible positions of the destination $n_d$ during the entire lifetime of the RREQ generated by $n_s$ at time ts, is termed as the Destination Embedding Circle (DEC) as observed by $n_i$. Its center is $(x_d(t1), y_d(t1))$ and radius is $\{v_{max}(d) \times ((t\text{-}t1)+\tau\text{-}(t\text{-}ts))\}$ i.e., $\{v_{max}(d) \times (\tau\text{-}(t1\text{-}ts))\}$.

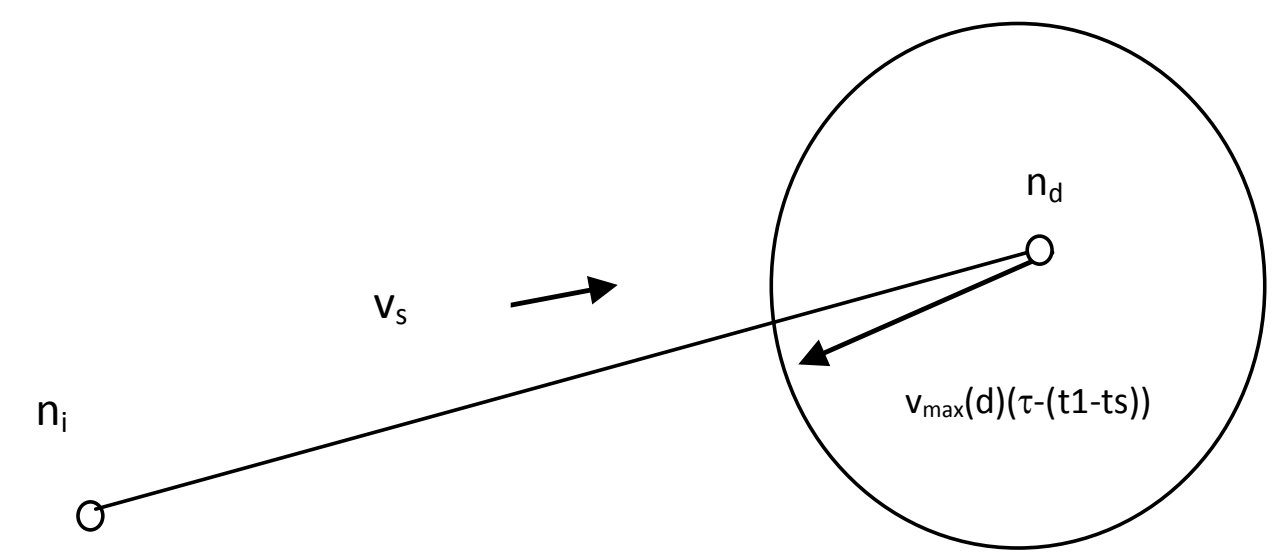

**Figure 1.** Demonstration of DEC.

$n_i$ will send RREQ packet to $n_j$ if any wireless signal transmitted by $n_j$ at time t can reach the nearest point on the DEC as observed by $n_i$ at time t1, within the time interval $(\tau\text{-}(t1\text{-}ts))$. The distance that can be travelled by the wireless signal within the time interval $(\tau\text{-}(t1\text{-}ts))$ is given by $(v_s \times (\tau\text{-}(t1\text{-}ts)))$. This distance should not be lesser than the distance of $n_j$ from the nearest point on the DEC as observed by $n_i$ at time t1, for receiving the RREQ from $n_i$. The situation can be depicted from Figure 1.

### 2.2 Node Information and Scheduling in FSRR

The technique FSRR requires that each node maintains a cache of route-request packets it has forwarded for other nodes, recently. The fields of the cache are,

- Identification number of source $(n_s)$
- Number of route-requests of that source transmitted so far $(R_s)$
- Timestamp of forwarding the first route-request of the source

FSRR gives priority to the route-requests destined towards the nodes with known recent location. Order of these packets is determined by a fuzzy controller named

Scheduler. Scheduler accepts the input parameters named Recent Quotient (RQ) and Position Quotient (PQ). RQ is the output parameter of Time-efficiency Fuzzy Controller and PQ is the output of Position-efficiency Fuzzy Controller. Input parameters of Time-efficiency Fuzzy Controller are RTR (Recent Timestamp Ratio), AST (Advancement Scope in Time) and CDHT (Chance of Discovering Higher Timestamp). On the other hand, the input parameters of Position-efficiency fuzzy controller are DECR (Destination Embedded Circle Ratio) and DR (Distance Ratio). Block diagram of the overall system is shown in Figure 2.

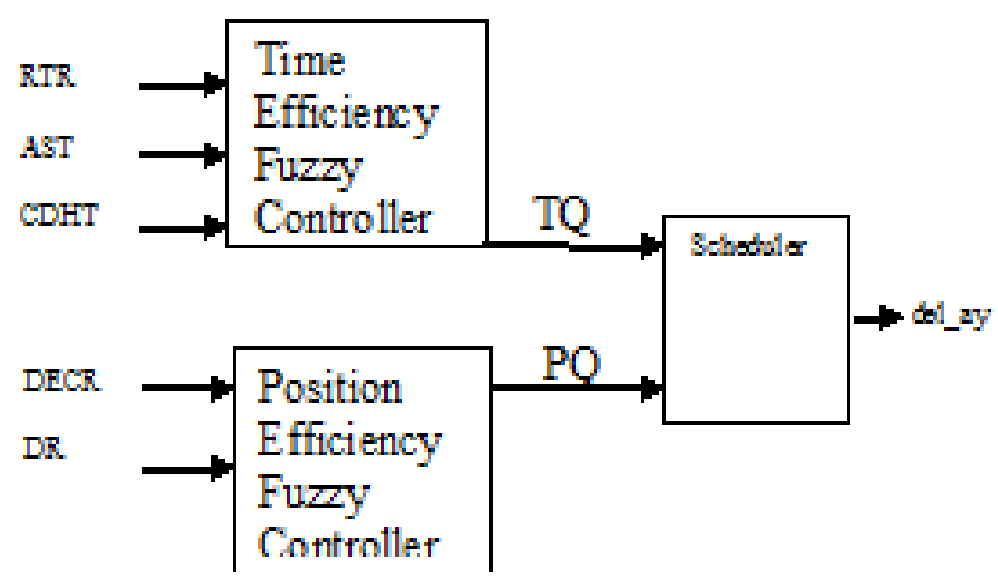

**Figure 2.** Block diagram of FSRR.

The route-requests to the destinations with unknown recent location are arranged in First-Come-First-Served order among themselves.

The above-mentioned fuzzy controllers work as per the following heuristics:

- If the embedding circle radius is small and distance between the current router and the last known location of the destination, is small then prioritizing the route-request will greatly reduce the delay in route discovery.
- If the timestamp of the most recent known location of the destination is very close to the current time, then that will reduce the unnecessary route-request packets in the network reducing the message contention and collision as well as the unnecessary delay.
- If most of the downlink neighbors of the current router have more recent record of communicating with the same destination, then the DEC predicted by those downlink neighbors will be more accurate than the current router and also the size of those DECs will be smaller than the same produced by the current router. This generates the scope of advancement in terms of temporal efficiency and also increases the chance of discovering higher i.e., better timestamps.

## 3. Design of Time Efficiency Fuzzy Controller

### 3.1 Input Parameters of Time-efficiency Fuzzy Controller

The input parameters are described below:

- $RTR_{i,d}(t) = tmr / t$ (1)

tmr is the timestamp at which the location of the destination was noted last. T is the current time. As per the formulation in Equation (1), if recent timestamp ratio is close to 1, then it will contribute to reduce the radius of the destination embedding circle and also inculcate accuracy in computing the region within which the route-requests will be transmitted.

$$AST_{i,d}(t) = \begin{cases} MAX \quad TB_{i,j,d}(t) \, if \, CLS_{i,d}(t) \, is \, non-null \\ \quad n_j \in CLS_{i,d}(t) \\ 0 \quad Otherwise \end{cases} \qquad (2)$$

For each $n_j \in N_i(t)$, let $\varsigma_{j,d}(t)$ be the largest timestamp of communication between $n_j$ and $n_d$ so far, where t is the current timestamp. If $\varsigma_{j,d}(t) > \varsigma_{i,d}(t)$ then $n_j$ is supposed to produce benefit in terms of recent timestamp than $n_i$. The time benefit $TB_{i,j,d}$ (t) produced by $n_j$ over $n_i$ w.r.t. $n_d$ at time t, is formulated in Equation (3).

$$TB_{i,j,d}(t) = 1 - \varsigma_{i,d}(t) / \varsigma_{j,d}(t) \qquad (3)$$

$CLS_{i,d}(t)$ is the set of those downlink neighbors $n_j$ s.t. $TB_{i,j,d}(t) \geq 0$. If it is null, then it indicates that there is no scope of advancement in terms of timestamp in the 1-hop downlink neighborhood of the current router $n_i$. Advancement in timestamp will definitely reduce the value of ($\tau$-(t1-ts)) and therefore reduce the radius as well as area of the destination embedding circle. So, higher the value of AST, lesser will be the radius of DEC.

$$CDHT_{i,d}(t) = F1_{i,d}(t) \times \{F2_{i,d}(t) / |CLS_{i,d}(t)|\}^{0.5} \qquad (4)$$

$$F1_{i,d}(t) = |CLS_{i,d}(t)| / |N_i(t)|$$

$$F2_{i,d}(t) = \Sigma(\varsigma_{j,d}(t) - AST_MN_{i,d}(t)) / (F2_1_{i,d}(t)+1)$$
$$n_j \in CLS_{i,d}(t)$$

$$F2_1_{i,d}(t) = (AST_{i,d}(t) - AST_MN_{i,d}(t))$$

Where $AST_MN_{i,d}(t) = MIN \;\; TB_{i,j,d}(t)$
$$n_j \in CLS_{i,d}(t)$$

The formulation is based on the idea that among the downlink neighbors that are producing more recent record of communication with the destination than the current router, the ones which are not producing the maximum timestamp but a timestamp close to the maximum, some of their successors might produce the largest timestamp of communication later on, during the traversal of route request packets from source to the DEC of destination.

$F1_{i,d}(t)$ ranges between 0 and 1 whereas $F2_{i,d}(t)$ ranges between 0 and $\{(AST_{i,d}(t) - AST_MN_{i,d}(t))/ (AST_{i,d}(t) - AST_MN_{i,d}(t)+1)\}$. $F1_{i,d}(t)$ deals with the fact that if a huge number of downlink neighbors of the current router generate higher timestamp of communication for the destination than the current router $n_i$, then the chance of discovering more better timestamps greatly increase. Those downlink neighbors again have their successors which may generate higher timestamps of communication with the same destination, than their predecessors. Even the 1-hop downlink neighbor of ni that is not producing the best timestamp at this point of time, some of its successor might produce the best one at a later stage. This is taken care of by $F2_{i,d}(t)$.

### 3.2 Rule Bases of Time-Efficiency Fuzzy Controller

RTR and AST range between 0 and 1. They are uniformly divided into crisp ranges. 0-0.25 is indicated as fuzzy premise variable a, 0.25-0.50 as b, 0.50-0.75 as c and 0.75-1.00 as d. CDHT range between 0 and MAXVAL where $MAXVAL = \{(AST_{i,d}(t) - AST_MN_{i,d}(t))/ (AST_{i,d}(t) - AST_MN_{i,d}(t)+1)\}^{0.5}$. It is divided into 0 – MAXVAL/4 (fuzzy variable a), MAXVAL/4 – MAXVAL/2 (fuzzy variable b), MAXVAL/2 – (3× MAXVAL)/4 (fuzzy variable c) and (3× MAXVAL)/4 to MAXVAL (fuzzy variable d).

Table 1 presents the fuzzy composition of RTR and AST producing temporary output temp1. The best combination is when RTR = d and AST = d. Temp1 is combined with CDHT generating output TQ. Here also the best combination is temp1 = d and CDHT = d.

**Table 1.** Fuzzy combination of RTR and AST producing temp1

| RTR → AST↓ | a | b | c | d |
|---|---|---|---|---|
| a | a | b | b | b |
| b | b | b | b | c |
| c | b | b | c | d |
| d | b | c | d | d |

**Table 2.** Fuzzy combination of temp1 and CDHT producing TQ

| Temp1 → CDHT↓ | a | b | c | d |
|---|---|---|---|---|
| a | a | a | b | c |
| b | a | b | c | c |
| c | a | b | c | d |
| d | b | c | d | d |

## 4. Design of Position Efficiency Fuzzy Controller

The input parameters are DECR and DR. These are mathematically formulated in subsection A and combined in fuzzy rule bases in subsection B in this section.

### 4.1 Input Parameters of Position-efficiency Fuzzy Controller

The input parameters are described below:

$$DECR_i = v_{max}(d) / V_MAX \quad (5)$$

V_MAX is the maximum of velocities of all nodes in the network. DECR ranges between 0 and 1. If it is small, then the embedding circle radius is also small. This will contribute to reduce the number of route-request packets.

- $DR_{i,d}(t) = \sqrt{\{(x_i(t) - x_d(t))^2+(y_i(t) - y_d(t))^2\}} \quad (6)$

It specifies the Cartesian distance between the two nodes $n_i$ and $n_d$ at time t. DR also ranges between 0 and 1. Values close to 0 reduce the number of required route-request packets.

### 4.2 Rule Bases of Position-Efficiency Fuzzy Controller

All the input parameters range between 0 and 1. They are uniformly divided into crisp ranges. 0-0.25 is indicated as fuzzy premise variable a, 0.25-0.50 as b, 0.50-0.75 as c and 0.75-1.00 as d. Table 3 combines DECR and DR producing the temporary output PQ. PQ gets the best combination when both DECR and DR are small i.e., both are a. Similarly, the worst combination is DECR = d and DR = d.

**Table 3.** Fuzzy combination of DECR and DR producing PQ

| DECR → DR↓ | a | b | c | d |
|---|---|---|---|---|
| a | d | d | c | b |
| b | d | c | c | b |
| c | c | c | b | a |
| d | b | b | b | a |

### 4.3 Design of Scheduler

The input parameters are TQ and PQ. These are already mathematically formulated in Sections 3 and 4 and combined in fuzzy rule base below.

All the input parameters range between 0 and 1. They are uniformly divided into crisp ranges. 0-0.25 is indicated as fuzzy premise variable a, 0.25-0.50 as b, 0.50-0.75 as c and 0.75-1.00 as d. Table 4 combines TQ and PQ producing del_ay. Del_ay gets the best combination when both TQ and PQ are large i.e., both are d. Similarly, the worst combination is TQ = a and PQ = a.

**Table 4.** Fuzzy combination of TQ and PQ producing del_ay

| TQ → PQ↓ | a | b | c | d |
|---|---|---|---|---|
| a | a | a | b | c |
| b | a | a | c | d |
| c | b | b | c | d |
| d | b | b | d | d |

The route-requesting node for which del_ay = d, gets served first. If multiple such nodes requests are waiting in the message queue, then they are served in FCFS order among themselves.

## 5. Simulation

Simulation of the mobile network has been carried out using ns-2[15] simulator on 800 MHz Pentium IV processor, 40 GB hard disk capacity and Red Hat Linux version 6.2 operating system. Graphs appear in Figures 2 and 3 showing emphatic improvements in favor of FSRR. Number of nodes has been taken as 20, 50, 100, 150 and 200 in different independent simulation studies. Speed of a node is chosen as 5 m/s, 10 m/s, 25 m/s, 35 m/s and 50 m/s in different simulation runs. Transmission range varied between 10 m and 50 m. Used network area is 500 m ×500 m. Used traffic type is constant bit rate. Mobility models used in various runs are random waypoint, random walk and Gaussian. Performance of the protocols AODV, ABR and FAIR are compared with their FSRR embedded versions FSRR-AODV, FSRR-ABR and FSRR-FAIR respectively. In order to maintain uniformity of the implementation platform, we have used ns-2 simulator for all the above-mentioned communication protocols. The only relevant simulation matrix is per node waiting time in message queue per router, in seconds, for tracking destination (total waiting time in message queue in different communication sessions /(total number of nodes × total number of routers)). Simulation time was 1000 sec. for each run.

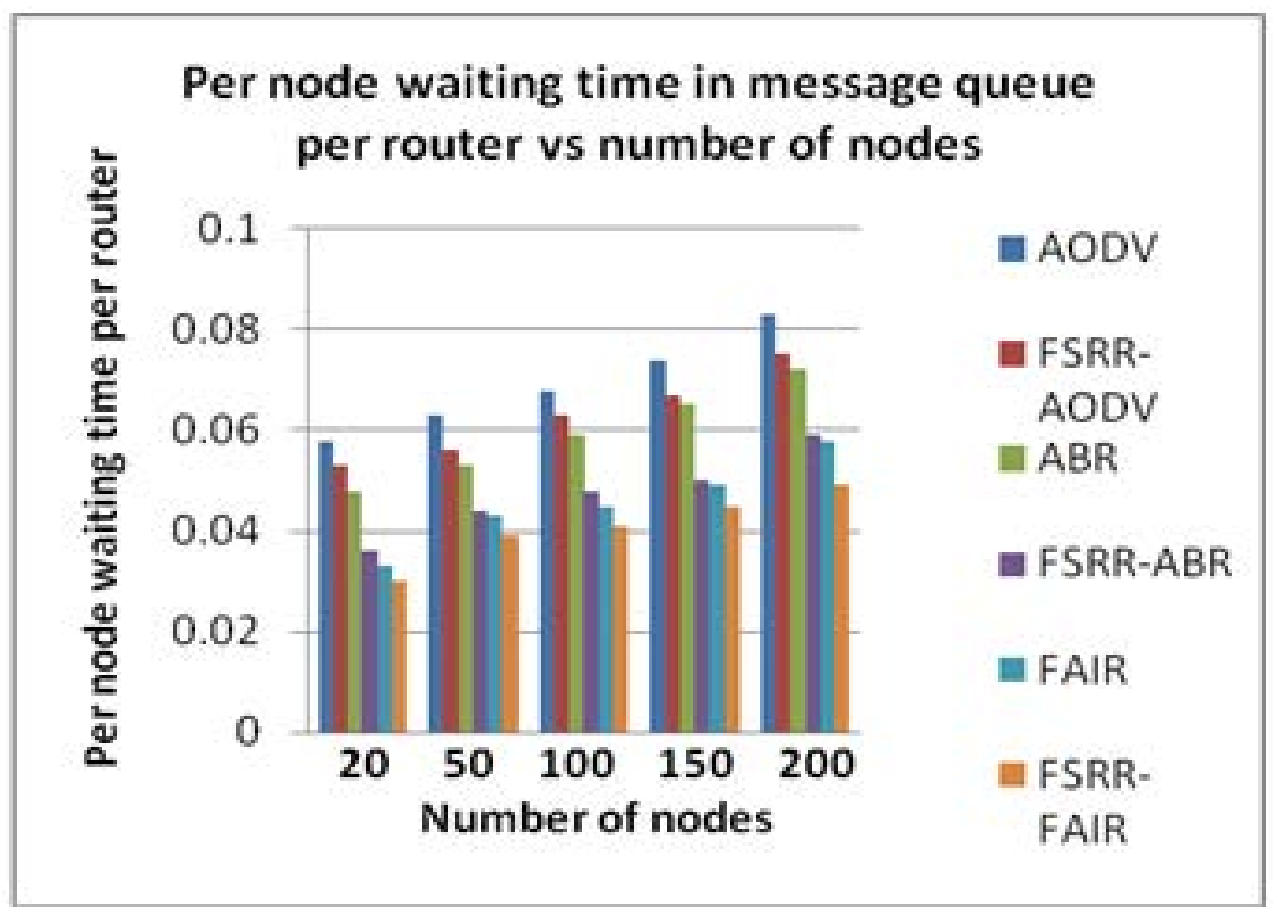

**Figure 3.** Graphical illustration of per node waiting time in message queue per router vs number of nodes (maximum node velocity is 25 m/s).

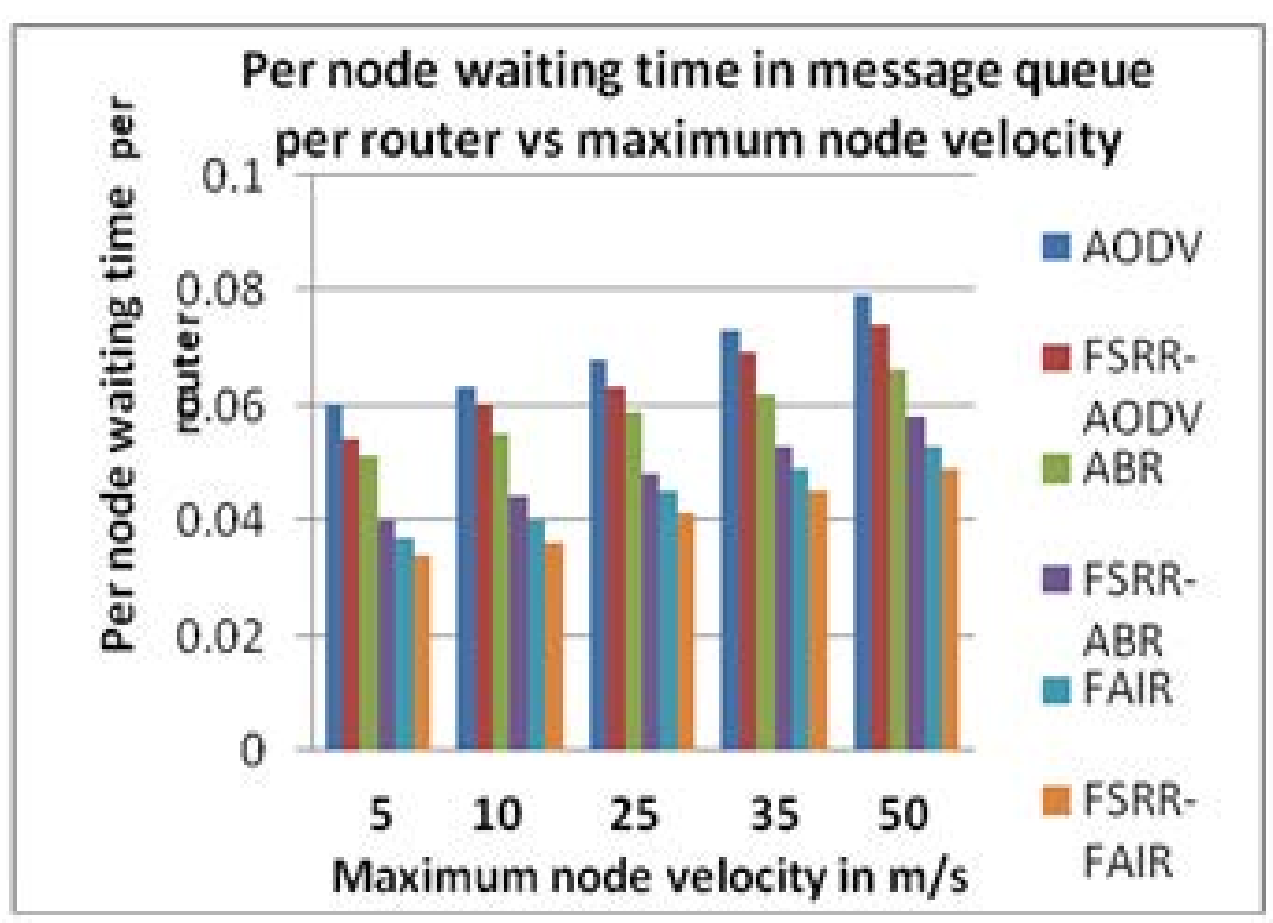

**Figure 4.** Graphical illustration of per node waiting time in message queue per router vs maximum node velocity (number of nodes is 100).

Figure 3 corresponds to the result when number of nodes varies between 20, 50, 100, 150, 200 and the maximum node speed is 25 m/s, whereas in Figure 4, maximum node velocity varies between 5 m/s, 10 m/s, 25 m/s, 35 m/s and 50 m/s, and number of nodes is kept constant at 100.

### 5.1 Experimental Results

As the number of nodes continues to increase, more route request packets arrive at routers. If the scheduling method is not efficient, a lot of delay will be unnecessarily produced by the system where the average waiting time will increase. Figure 3 shows that as the numbers of nodes goes on increasing, FSRR embedded versions of protocols produce much lesser average waiting time simply because it processes the route-requests with close destinations before the others at higher distances. Also it is a fact that the estimated locations of the destination should be accurate.

Figure 4 deals with the average waiting time versus the maximum node velocity. As the maximum node velocity increases, the links tend to break frequently in general. That will give rise to the injection of a huge number of route-request packets in order to repair the broken links. As a result, more route-requests will arrive at the routers. In this situation, FSRR embedded protocols efficiently schedule the route-requests and decrease the average waiting time to find the destinations.

## 6. Conclusion

Route discovery is a very important part in communication in ad hoc networks. If the average waiting time of route-request packets is decreased then automatically the time required to track the destination nodes will also decrease. FSRR is based on the concept that route-requests towards close destinations should be given priority. Great improvements are produced in FSRR embedded routing protocols than their ordinary versions.

## 7. References

1. Perkind CE, Bhagat P. Highly dynamic Destination Sequenced Distance Vector routing (DSDV) for mobile computers. Computer Communications Review. 1994; 24(4):234–44.
2. Murphy S, Garcia-Luna-Aceves JJ. An efficient routing protocol for wireless networks. ACM Mobile Networks and Applications Journal. 1996 Nov; 183–97. Available from: http://citeseer.nj.nec.com/10238.html
3. Chen TW, Gerla M. Global state routing: A new routing scheme for wireless ad hoc networks. IEEE Conference on Information, Communication and Control; 1998.
4. Chiang CC, et al. Routing in clustered multi-hop wireless networks with fading channel. IEEE Conference on Innovative Systems; Singapore. 1997.
5. Broch J, Maltz D, Johnson D, Hu Y, Jetcheva J. A performance comparison of multi-hop wireless ad hoc network routing protocols. Proceedings of the 4th Annual ACM/IEEE International Conference on Mobile Computing and Networking (MobiCom '98); Dallas, Texas, USA. 1998 Aug.
6. Kodole A, Agarkar PM. A survey of routing protocols in mobile ad hoc networks. Multi-Disciplinary Journal of Research in Engineering and Tech. 2015; 2(1):336–41.
7. Perkins CE, Royer EM. Ad hoc on-demand distance vector routing. IEEE Workshop on Mobile Computing Systems and Applications; 1999.
8. Brown TX, Babow HN, Zhang Q. Maximum flow life curve for wireless ad hoc networks. ACM Symposium on Mobile Ad Hoc Networking and Computing; USA. 2001.
9. Toh CK, Bhagwat P. Associativity based routing for mobile ad hoc networks. Wireless Personal Communications. 1997; 4(2):1–36.
10. Banerjee A, Dutta P. Fuzzy-controlled Adaptive and Intelligent Route selection (FAIR) in ad hoc networks. European Journal of Scientific Research. 2010; 45(3):367–82.
11. Victoria DRS, Kumar SS. Efficient bandwidth allocation for packet scheduling. International Journal of Future Computing and Communications. 2012 Dec; 1(4).
12. Annadurai C. Review of packet scheduling algorithms in mobile ad hoc networks. International Journal of Computer Applications. 2011 Fed; 15(1).
13. Banerjee A, Dutta P. Delay-efficient Energy and Velocity Conscious Non-Preemptive Scheduler (DEV-NS) for mobile ad hoc networks. International Journal of Advanced Networking and Applications. 2014; 5(4).
14. Parasher R, Rathi Y. A-AODV: A modern routing algorithm for mobile ad hoc networks. International Research Journal of Engg and Technology. 2015; 2(1).
15. Available from: http://www.isi.edu/nsnam/ns/tutorial/